\documentclass[a4paper]{jpconf}

\let\TW=\textwidth
\def\RE{\ensuremath{\mathrm{Re}}}
\def\RT{\ensuremath{\RE_{\rm t}}}
\def\RG{\ensuremath{\RE_{\rm g}}}

\usepackage{amsmath,amssymb}
\usepackage{graphicx}
\usepackage{natbib}
\begin{document}
\title{On the decay of turbulence in plane Couette flow}

\author{Paul Manneville}

\address{Hydrodynamics Laboratory, \'Ecole Polytechnique, Palaiseau, France}

\ead{paul.manneville@polytechnique.edu}

\begin{abstract}
The decay of turbulent and laminar oblique bands in the lower transitional range of plane Couette flow is studied by means of direct numerical simulations of the Navier--Stokes equations. We consider systems that are extended enough for several bands to exist, thanks to mild wall-normal under-resolution considered as a consistent and well-validated modelling strategy. We point out a two-stage process involving the rupture of a band followed by a slow regression of the fragments left. Previous approaches to turbulence decay in wall-bounded flows making use of the chaotic transient paradigm are reinterpreted within a spatiotemporal perspective in terms of large deviations of an underlying stochastic process. 
\end{abstract}

\section{Context}
Difficulties in understanding the transition to turbulence in wall-bounded flow arise from its hysteretic character linked to the fact that the nontrivial turbulent regime stands at a distance from the laminar regime and coexists with it in some range of Reynolds numbers \RE. 
In plane Couette flow -- the flow between counter-translating parallel plane plates (speed $\pm U$, gap $2h$, PCF for short) --  the laminar (trivial) base state is linearly stable for all Reynolds numbers $\RE=Uh/\nu$ and the turbulent state is uniform above $\RT\approx410$.
Below $\RG\approx325$, laminar flow prevails in the long-time limit.
Between \RG\ and \RT, the nontrivial state displays a pattern of alternately turbulent and laminar oblique bands, at least in systems with in-plane dimensions very large when compared to the gap \citep{Petal03}.
The decay of turbulence in the vicinity of \RG\ has mostly been theoretically analysed within the framework of {\it low dimensional dynamical systems\/}, using periodic domains called {\it minimal flow units\/} (MFUs) with periodic boundary conditions at distances of the order of the gap between the plates \citep{Wa97}.
This approach helps us to understand the general mechanisms sustaining the turbulent regime  and permits us to use the concepts and tools of {\it deterministic chaos\/} theory.
Within this framework, decay is described in terms of {\it chaotic transients\/} characterised by lifetimes with exponentially decreasing  probability distributions \citep{Eetal08}.
However, from a practical point of view, the {\it extended\/} geometries of interest in experiments rather appeal to concepts of {\it pattern formation\/} and {\it spatiotemporal chaos\/}.
Here we focus on the lower end of the transitional range around \RG, on how the bands become fragmented into turbulent patches that finally decay \citep{Ma11}, and on the extent to which this can change our views on turbulence breakdown, in contrast with what was put forward at the MFU scale.

\section{Results}

We perform direct numerical simulations using Gibson's public domain code {\sc ChannelFlow}   (www.channelflow.org).
This pseudo-spectral Fourier$\,\times\,$Chebyshev$\,\times\,$Fourier code is used at voluntarily degraded resolution, understood as a consistent modelling strategy \citep{MaRo10}.
\begin{figure}
\begin{center}
\includegraphics[height=0.33\TW,clip]{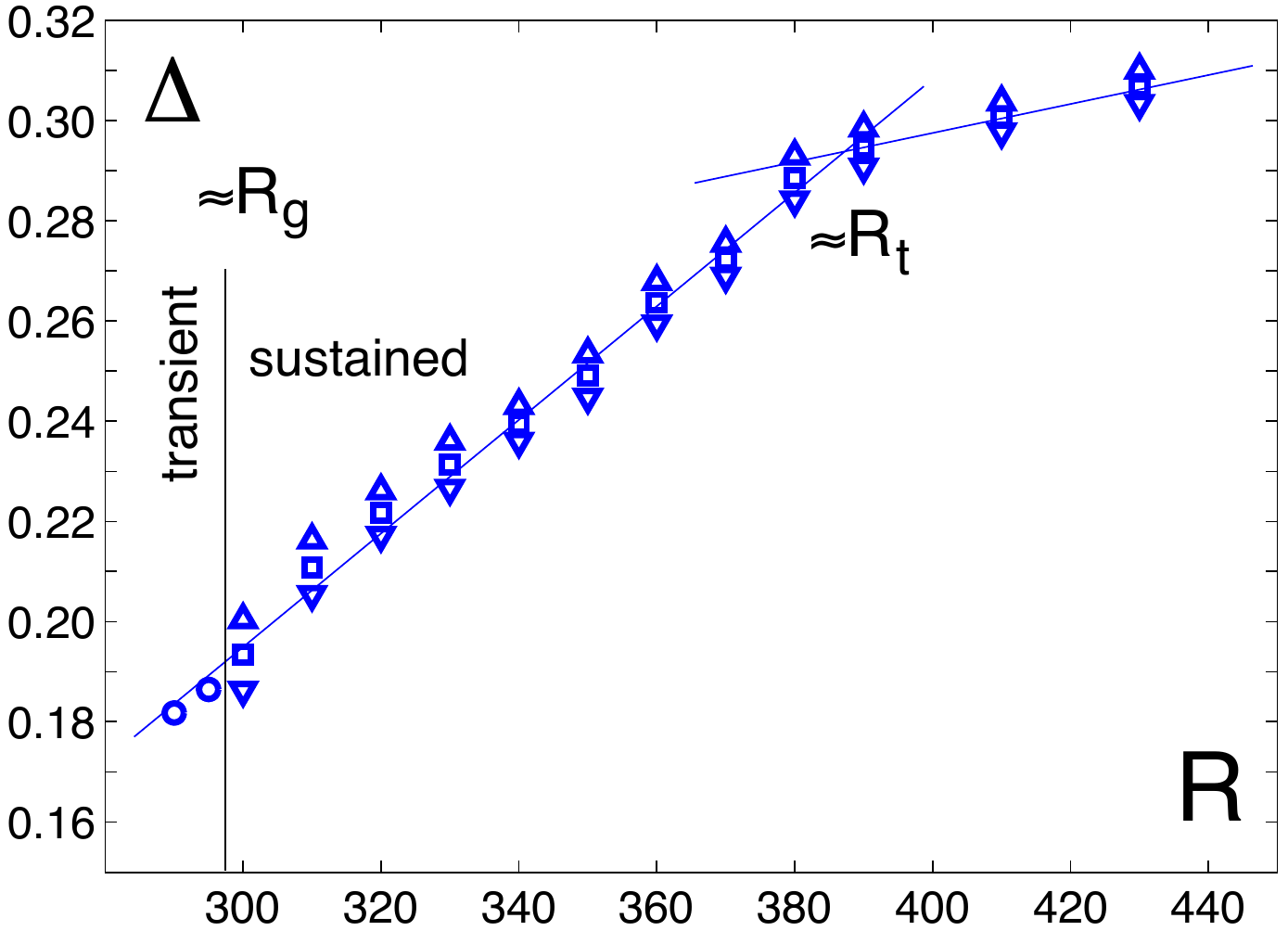}
\hskip0.5em
\includegraphics[height=0.33\TW,clip]{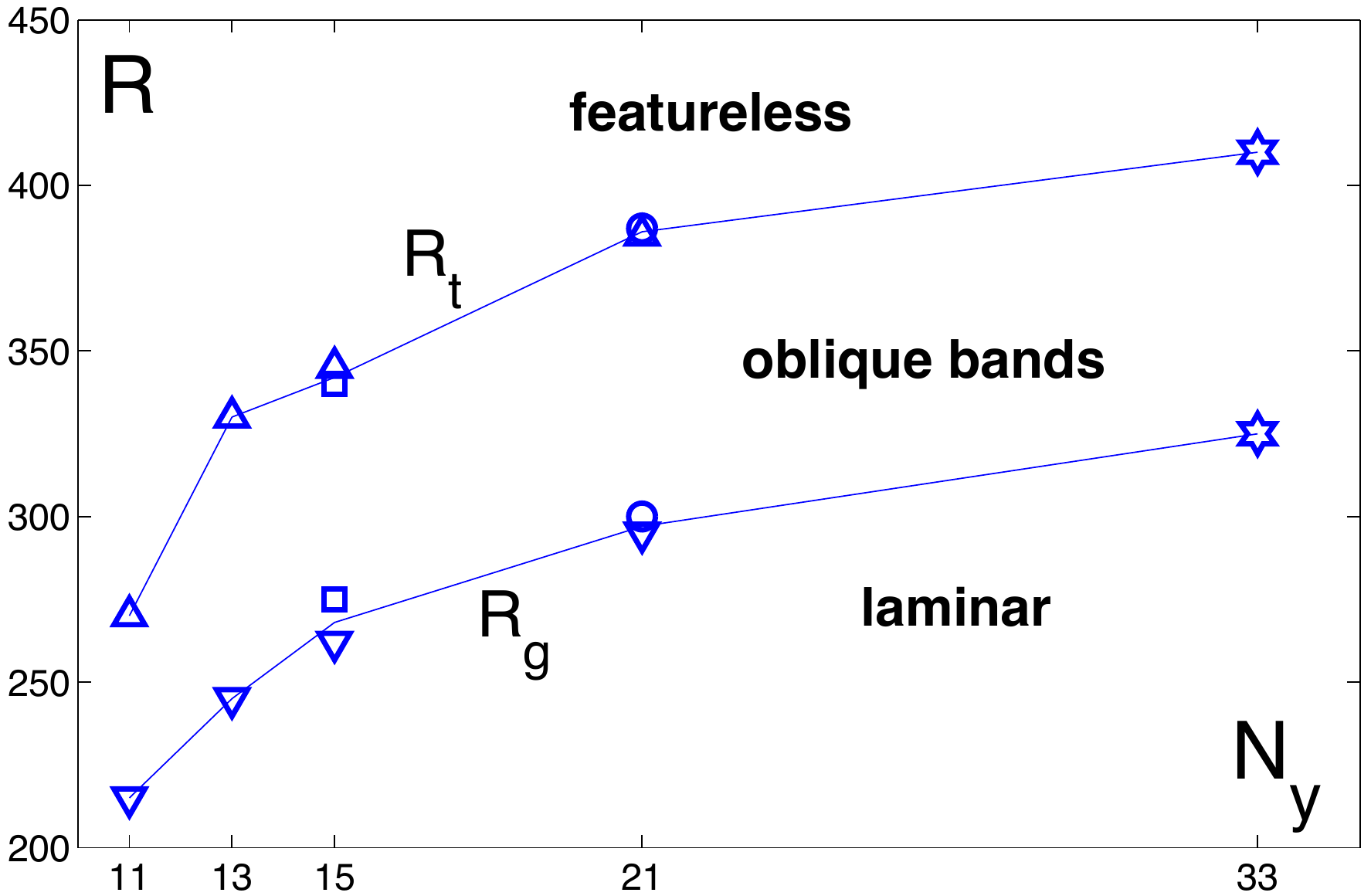}
\vspace*{-0.5\baselineskip}

\end{center}
\caption{\label{fig0} Variation of the time-average of the distance  $\Delta$ to laminar flow as a function of $R$
for $N_y=21$ with $(L_x,L_z)=(110,84)$ and  $(N_x,N_z)=(440,336)$, see \citep{MaRo10}.
Sustained bands are conspicuous between $\RG$ below which they inevitably decay in the long term, and $\RT$ above which uniform turbulence is obtained.
Similar experiments for different in-plane resolutions and domain sizes helped us to build the diagram on the right displaying thresholds $R_{\rm g}$ and $R_{\rm t}$ as functions of the number $N_y$ of Chebyshev polynomials. Marked by stars, the values obtained for $N_y=33$ \citep{Detal10} cannot be distinguished from experimental findings \cite{Petal03}. Below $N_y=11$, no bands are found like in \cite{Ma09} resting on the low order Galerkin modelling of \cite{LM07} . $N_y=15$ is found to be an acceptable compromise allowing the consideration of domains wide enough to accommodate several bands.}
\end{figure}
Preliminary work showed that the band regime is remarkably robust, see Fig.~\ref{fig0}. For systems with in-plane dimensions $L_x$ (streamwise) and $L_z$ (spanwise), the main features of transitional PCF are appropriately reproduced, including the pattern's wavelengths,  with $N_x=L_x$ and $N_z=3L_z$ collocation points, while taking  $N_y=15$ Chebyshev polynomials just yields a downward shift of $[\RG,\RT]$ of about 15\%  to  $\approx[275,350]$.
Using this optimal compromise between accuracy and computational load, we consider a pretty wide domain of size $L_x=432h$ and $L_z=256h$ with the power of a desktop computer only.

\begin{figure}
\begin{center}
\includegraphics[angle=90,width=0.18\TW,clip]{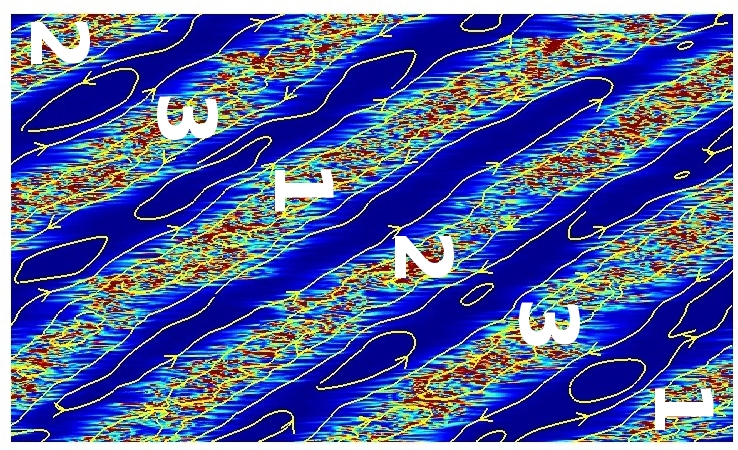}
\hskip0.5em
\includegraphics[angle=90,width=0.18\TW,clip]{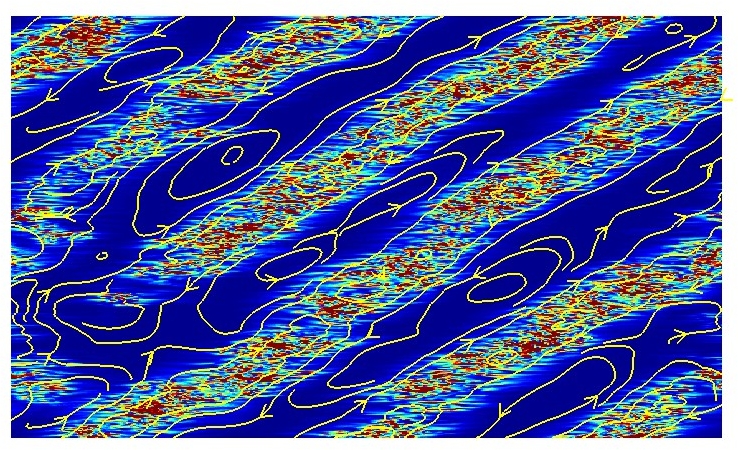}
\hskip0.5em
\includegraphics[angle=90,width=0.18\TW,clip]{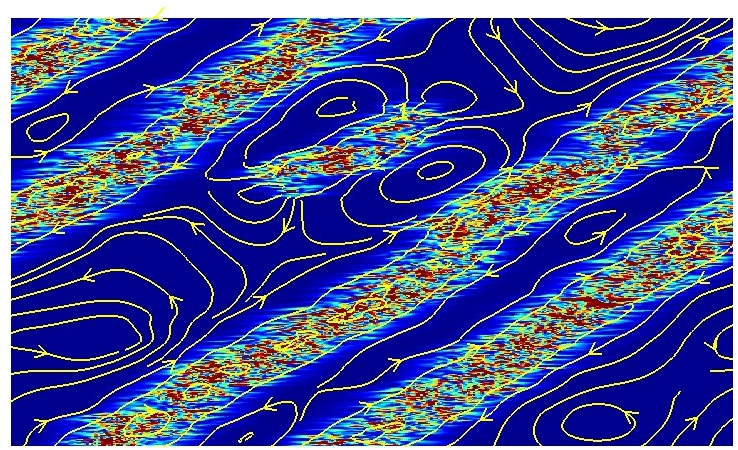}
\hskip0.5em
\includegraphics[angle=90,width=0.18\TW,clip]{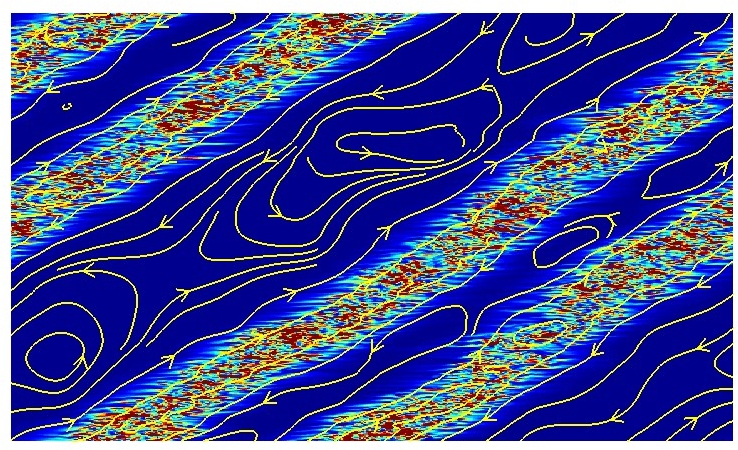}
\hskip0.5em
\includegraphics[angle=90,width=0.18\TW,clip]{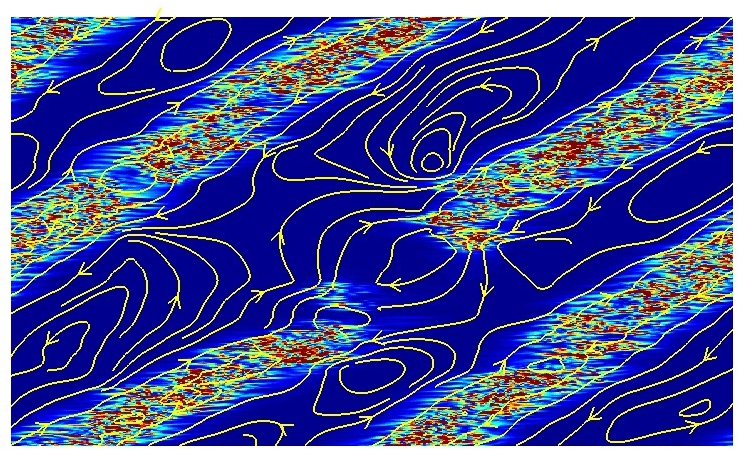}\\[2ex]
\includegraphics[angle=90,width=0.18\TW,clip]{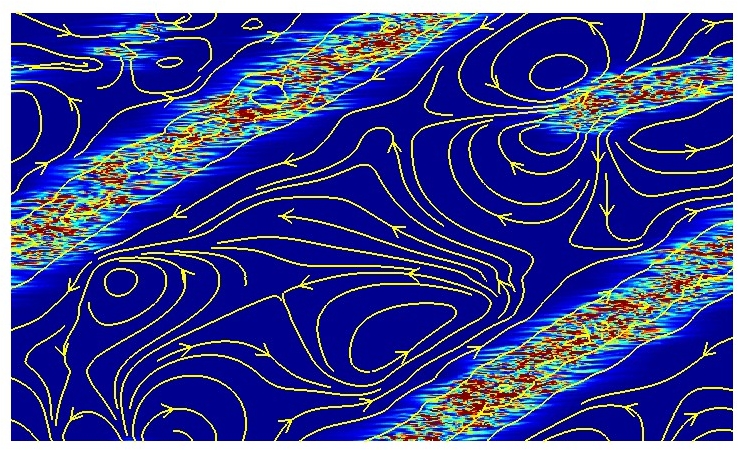}
\hskip0.5em
\includegraphics[angle=90,width=0.18\TW,clip]{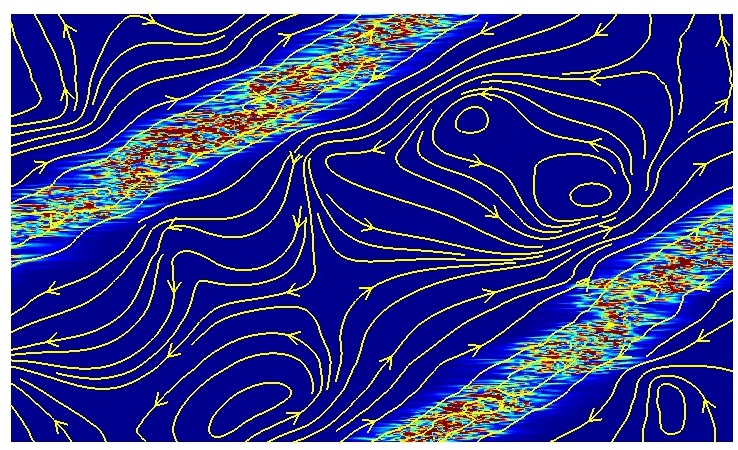}
\hskip0.5em
\includegraphics[angle=90,width=0.18\TW,clip]{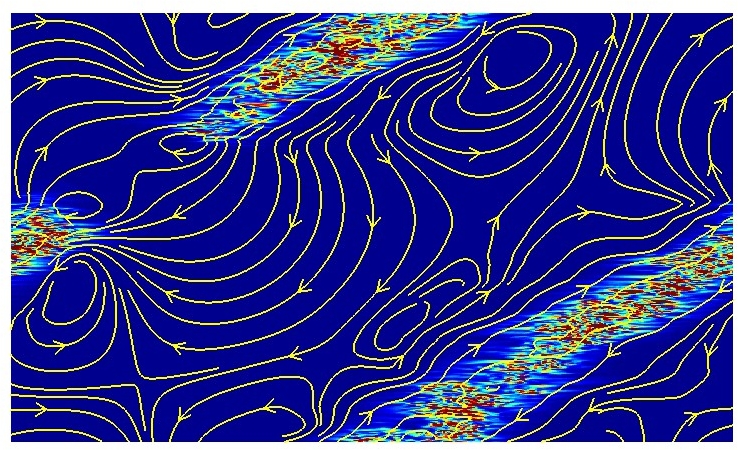}
\hskip0.5em
\includegraphics[angle=90,width=0.18\TW,clip]{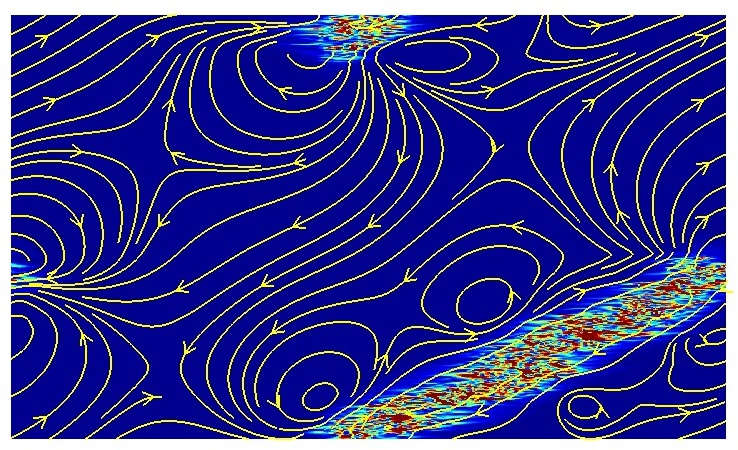}
\hskip0.5em
\includegraphics[angle=90,width=0.18\TW,clip]{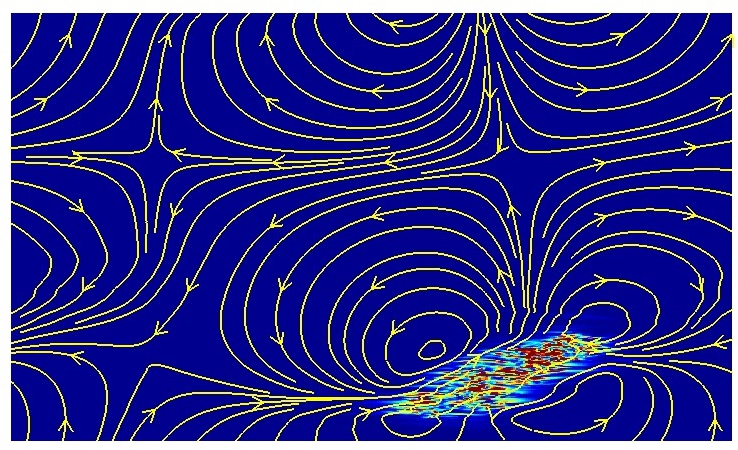}
\end{center}
\caption{Snapshots taken during the decay at $\RE=272.5$. From left to right and top to bottom.
$t=0$: initial state, three continuous bands (1,2,3); labelling takes periodic boundary conditions into account. $t=2250$: band 1 breaks. $t=5250$: band 1 recedes. $t=6000$: two continuous bands (2,3). $t=15750$: band 2 breaks. $t=18750$: band 2 recedes. $t=20250$: one continuous band (3). $t=21750$: band 3 breaks. $t=25500$ \& $26250$, band 3 recedes and decays.
Time is reset at the beginning of the experiment when $\RE$ is quenched from 275.0 to 272.5.
The streamwise direction is along the vertical.
Colour scale for local perturbation energy from blue for laminar to red for peaks of turbulence.
Streamlines of the large scale fow, conspicuous in the laminar regions, are indicated.
This flow is very weak except close to the turbulent patches.
Its skewed quadrupolar shape around the tips of turbulent band fragments is typical.\label{fig1}}
\end{figure}

Starting from a state in the featureless regime at $\RE=450\gg\RT$, the Reynolds number is decreased by steps down to the range of interest around $\RG\approx275$.
At $\RE=275$ a steady pattern with 3 oblique bands is obtained with streamwise period $\lambda_x=144h$ and spanwise period $\lambda_z=85.3h$, in reasonable agreement with experimental findings by \citet{Petal03}.
When \RE\ is decreased further, the pattern breaks and decays.
Figure~\ref{fig1} illustrates the different steps observed during the transient that starts from a state at $\RE=275$ upon setting $\RE=272.5$.

\begin{figure}
\centerline{\includegraphics[height=0.55\TW,clip]{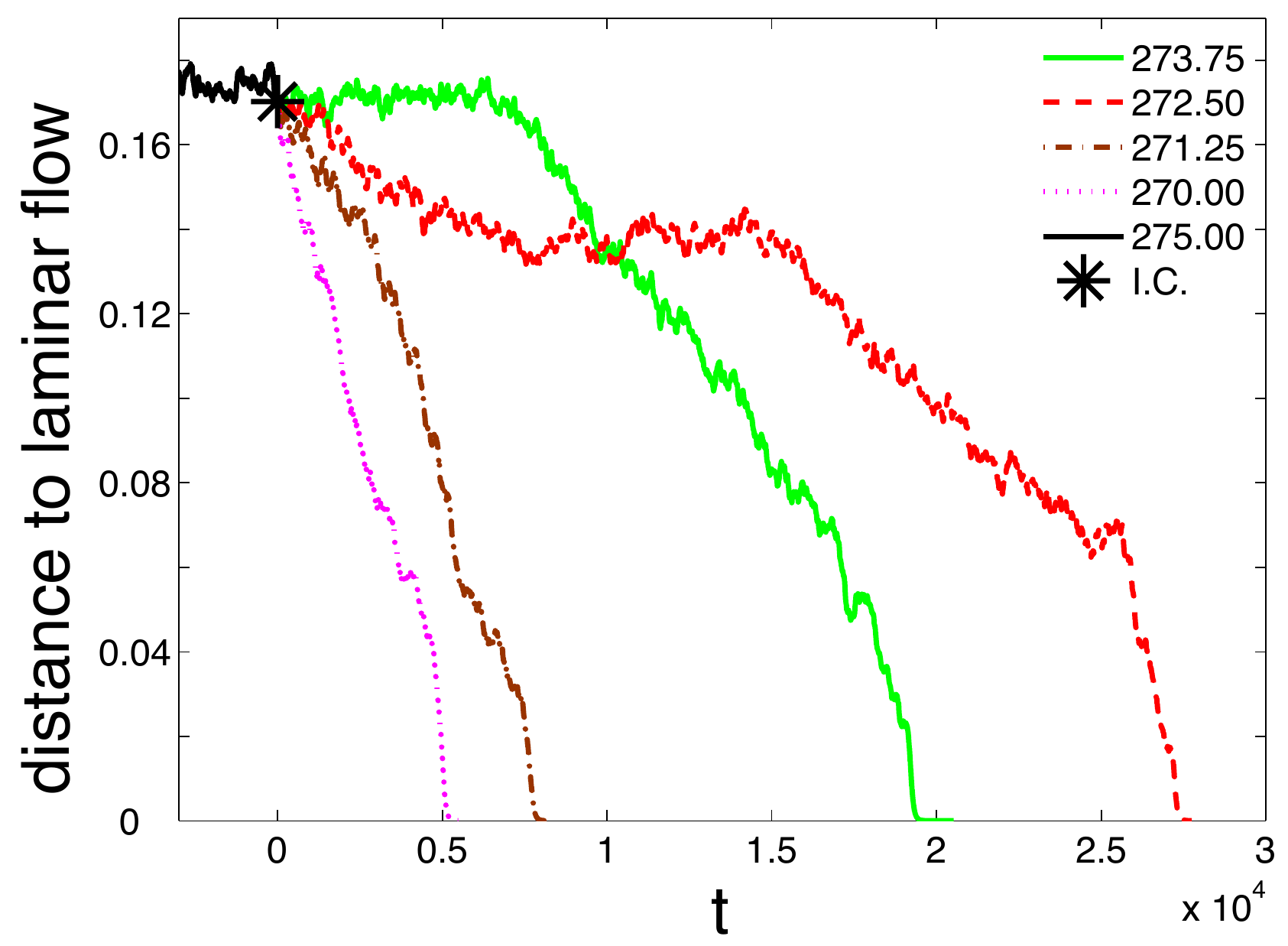}}
\caption{Time series of the distance to laminar flow during transients started from the same initial condition and different values of \RE\ in the wide system ($432\times256$).\label{fig2}}
\end{figure}

Two different processes are seen to take place: the breaking of a continuous band and the slow retraction of band fragments at roughly constant speed.
These processes are always present in the experiments that have been undertaken, some of which are featured by means of time series of the distance $\Delta$ to laminar flow in Figure~\ref{fig2}. 
Band breaking is crucial in controlling the decay. For example, the transient at $\RE=273.75$ is marked by the persistence of the 3-band state for about $7000\,h/U$ before two bands break nearly simultaneously (green full line) and recede. In contrast,  in the transient at $\RE=272.50$ pictured in Figure~\ref{fig1}, bands break one after the other and the 2-band state remains unbroken during about $8000\,h/U$ (red dashed line). The three bands break shortly after the beginning of the experiments at $\RE=271.25$ and $270$ (brown and magenta lines).

While band breaking leaves essentially no trace on the time series of $\Delta$ (Fig.~\ref{fig2}), the slow withdrawal of the band fragments corresponds to the linear decrease of this quantity which is roughly proportional to the cumulated length of the band fragments. The decrease is then easily understood as retraction at constant speed, itself a function of the number of fragments since retraction takes place at each end of a given fragment, statistically. This explains that the decay of turbulence is faster for $\RE=273.75$ with two simultaneously broken bands than for $\RE=272.5$ with only one band broken at a time. Both processes, the breaking of bands and the trimming of band fragments at their extremities are attributed to the chaotic dynamics at the MFU scale: Being transient in the range of \RE\ of interest, it implies a finite probability for turbulence to decay locally.

\begin{figure}
\centerline{\includegraphics[width=0.9\TW,clip]{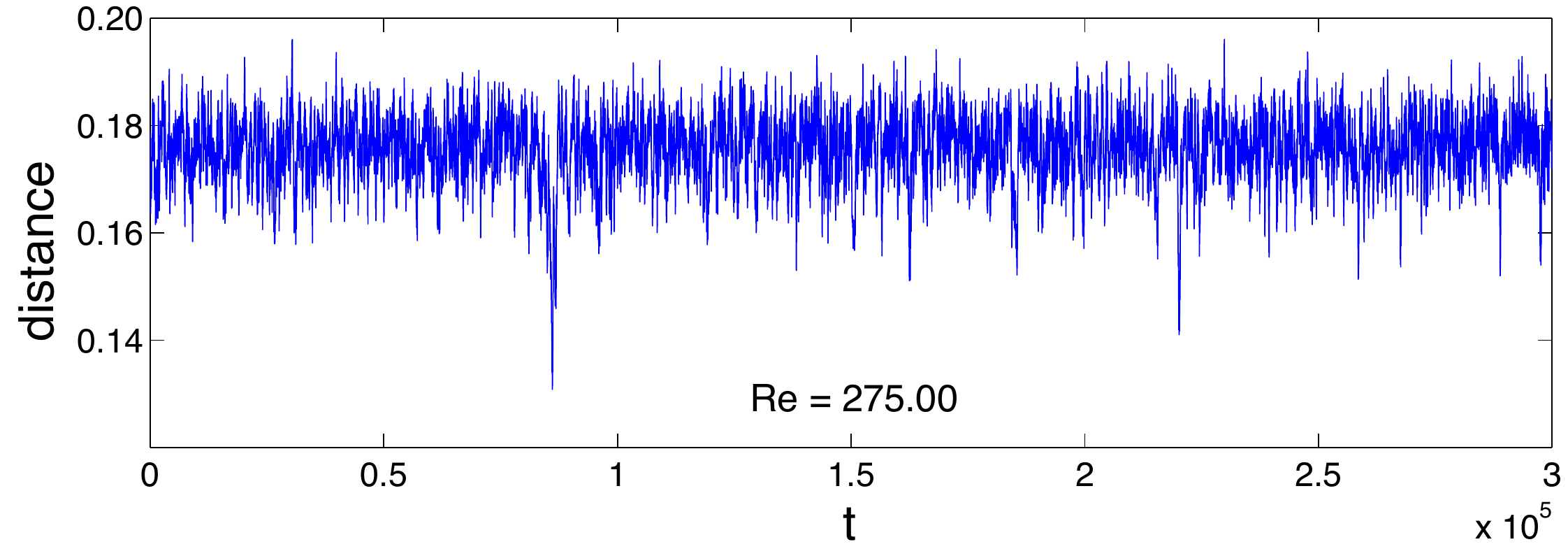}}
\caption{Time series of the distance to laminar flow for the sustained banded state at $\RE=275$ in the small system ($144\times84$).\label{fig3}}
\end{figure}

Band breaking happens when a turbulent band collapses in a domain of size at least as large as the band itself. This specific process has been studied in a smaller domain of size $144\!\times\!84$ able to contain just a single wavelength of the pattern \citep{RoMa11}. Figure~\ref{fig3} displays the time series of the distance to laminar flow for $\RE=275$ during an experiment of  very long duration ($3\!\times\!10^5\>h/U$) over which this specific  turbulent regime remains sustained.
Excursions toward low values of the distance, such as the one at $t\approx8500\,h/U$ are not sufficiently dangerous nor sufficiently frequent to trigger the decay of the system to laminar flow. By contrast, for $\RE<275$, decay is observed in finite duration simulations, owing to the more frequent occurrence of more dangerous excursions.

\begin{figure}
\centerline{\includegraphics[height=0.55\TW,clip]{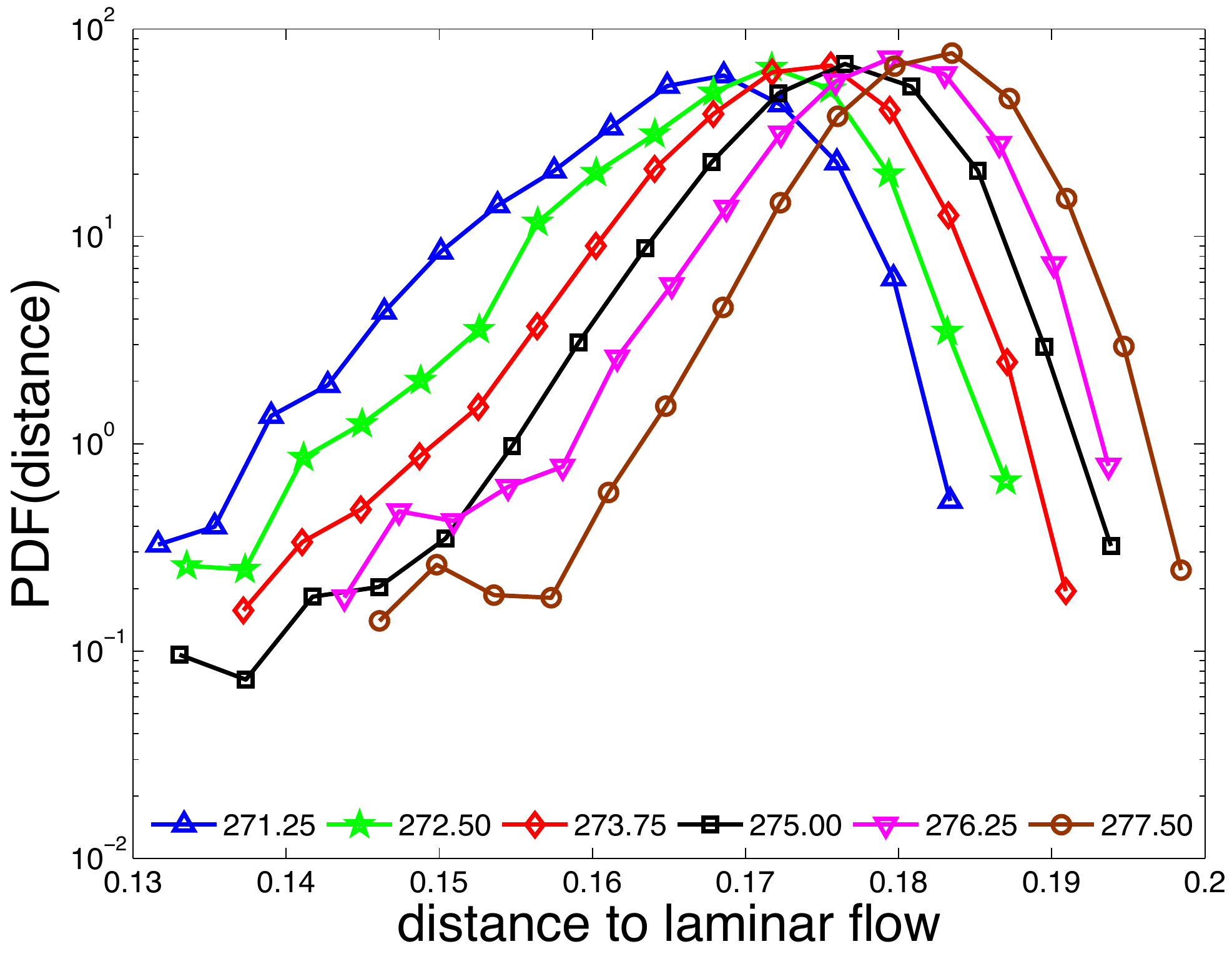}}
\caption{PDFs of the distance to laminar flow in the small system for different values of \RE.\label{fig4}}
\end{figure}

Probability density functions (PDF) of the instantaneous value of $\Delta$ are then constructed.
They are displayed in Figure~\ref{fig4} for different values of \RE, where it  can be seen that the most probable value slowly decreases with \RE\ but, more importantly, that the probability of low values, corresponding to the dangerous excursions, increases drastically in a narrow range of \RE.  (Irreversible decay takes place for distance $< 0.13$.)
Decay is not observed for $\RE\ge275$ for time series longer than $3\!\times\!10^5$, though the exponential  tail in the corresponding PDF suggest that this event could occur with very low probability, i.e. could be observable in much longer simulations only.

After a sufficiently wide gap has been nucleated, the steady retraction of the turbulent domains is then easily understood as a noisy propagation problem for a stable state (laminar) that gains over a metastable one (turbulent). This process can be understood in terms of \emph{spatiotemporal intermittency} below threshold \citep{CM87}. Care should be taken that the band fragments are surrounded by a large scale mean flow, which somehow complicates the dynamics.

\section{Discussion}
A detailed understanding of turbulence decay in systems of practical interest asks for going beyond the {\it temporal\/} framework underlying the chaotic transients paradigm \citep{Eetal08}.
Although it is essential to notice that the bands are observed in an range of Reynolds numbers where chaos is transient, it seems necessary to include \emph{spatiotemporal} effects. 
The breakdown of turbulent bands then appears to be governed by a stochastic process involving coupled units featuring small clusters of tightly coupled MFUs with chaotic dynamics.
Chaos at such scales and its possible local collapse are at the origin of the \emph{large deviations} able to open gaps in the turbulent bands and are also responsible for the subsequent slow withdrawal of band fragments.
These observations substantiate Pomeau's  conjecture  \citep{Po86} about the relevance of nucleation processes in first-order thermodynamic-like phase transitions on the one hand, and of  spatiotemporal intermittency as a  {\it directed-percolation\/} stochastic process \citep{Ki83} on the other hand, while showing that the complete hydrodynamic problem keeps some specificities since the role of large scale flow corrections within laminar domains and the origin of the bands are so far not elucidated.
\bigskip

\noindent \small {\it Acknowledgments.}  The help  of J.~Rolland for the implementation of {\sc ChannelFlow} and discussions with F. Daviaud, Y.~Duguet, J.~Philip, A. Prigent, J.~Rolland, and L.S. Tuckerman are deeply acknowledged. 

\bibliographystyle{jfm}
\bibliography{etc13}

\end{document}